\documentstyle[prl,multicol,epsfig,aps,amsmath,amssymb]{revtex}

\begin{document}

\draft

\title{Carbon Nanotubes as Schottky Barrier Transistors}

\author{S.~Heinze, J.~Tersoff$^*$, R.~Martel, V.~Derycke, J.~Appenzeller, and Ph.~Avouris$^+$}

\address{IBM Research Division, T.~J.~Watson Research Center, Yorktown Heights, New York 10598}

\date{\today, appears in Physical Review Letters.}

\maketitle

\begin{abstract}
We show that carbon nanotube transistors operate
as unconventional ``Schottky barrier transistors'',
in which transistor action occurs primarily by
varying the contact resistance rather than the
channel conductance.
Transistor characteristics are calculated for
both idealized and realistic geometries,
and scaling behavior is demonstrated.
Our results explain a variety of experimental
observations, including
the quite different effects of doping and adsorbed gases.
The electrode geometry is shown to be crucial for
good device performance.
\end{abstract}


\vspace{-.7cm}

\begin{multicols}{2}

\narrowtext

Carbon nanotubes show great promise for nanoscale
field-effect transistors (FETs) \cite{Tans98,Martel98}.
However, despite considerable progress in device
fabrication \cite{Derycke01,Martel01,Dekker01},
the theoretical understanding remains incomplete.
Initially, it was naturally assumed that the gate voltage
modified the nanotube (NT) conductance, in analogy with
the channel of an ordinary FET.
However, there has been increasing evidence
that Schottky barriers at the contacts may play
a central role \cite{Martel01,Dekker01,Johnson01,Derycke02}.
In fact, ordinary FETs require ohmic contacts for effective switching,
while reasonable workfunction estimates
suggest significant Schottky barriers at NT-metal
contacts \cite{Leonard00,Odintsov00}.

Here we show that, whenever there is a substantial
Schottky barrier (SB) at the contact, NT-FETs operate
as unconventional {\it Schottky barrier transistors},
in which switching occurs primarily by modulation of
the contact resistance rather than the channel conductance.
SB-FETs have already been considered as a
possible future silicon technology \cite{SB-FET},
because of their potential to operate at extremely
small dimensions.
We calculate the characteristics of such NT devices,
and show that they explain a number of key experimental
observations, including the effect of dopants,
and the non-ideal switching behavior.
In particular, we find that the effect of adsorbed gases
can be explained simply by their effect on workfunctions,
without any doping as has been generally assumed.
The geometry of the contact electrode plays a central role,
primarily by scaling the required gate voltage.
We predict that the device characteristics
can be {\it radically improved} by tailoring the contact geometry.

\begin{figure}
\begin{center}
\epsfig{file=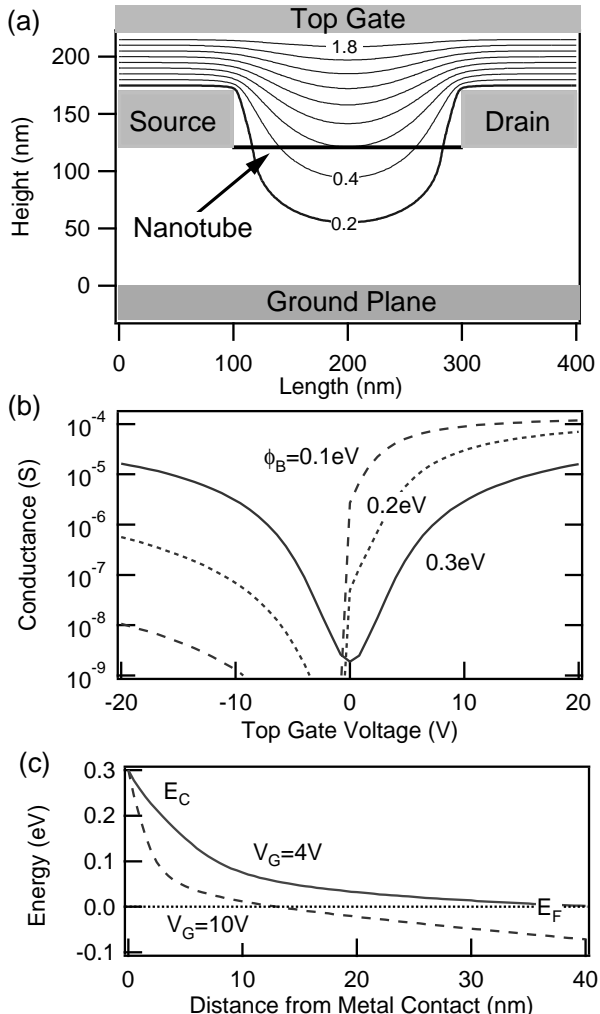,width=7.8cm,angle=0}
\caption{
\label{Fig:SB-variation}
Conductance for realistic FET geometries.
(a) Device geometry, with metal contacts on the left and right,
a ground plane, and a top gate.
Contour lines show the electrostatic potential for a top gate
voltage of 2~V.
(b) Corresponding conductance versus gate voltage
at room temperature, for different SBs.
The SB height for electrons is indicated for each curve.
(c) NT conduction-band energy near the contact,
for gate voltages of 4 and 10~V.}
\end{center}
\end{figure}

We begin by calculating the characteristics of the device
shown in Fig.~\ref{Fig:SB-variation}a,
consisting of a NT embedded in a dielectric
between a top gate and a ground plane, with two metal
electrodes as source and drain.
Such devices have recently been reported \cite{Schalom}.
We solve numerically for the self-consistent electrostatic potential
as in Ref.~\cite{Leonard99}, with the source, drain, and gate voltages
as boundary conditions.  For simplicity, the calculations in Fig.~1
use a local approximation to the electrostatic kernel for
charge on the NT \cite{local}.
This is justified below by
comparisons with a more complete calculation.

The current is then given by the Landauer formula,
\begin{equation}
I(V)=\frac{4e}{h} \int \left[ F(E)-F(E+eV_D) \right] P(E) dE ~.
\label{eq:current}
\end{equation}
Here $V_D$ is the drain voltage (with the source taken as ground),
and $F$ is the Fermi function.
We calculate the energy dependent transmission probability
$P(E)$ within the WKB approximation
for a semiconducting NT of bandgap 0.6~eV.
(WKB is relatively accurate when the tunneling resistance is high,
so it describes the overall ``turn-on" well.
However, it neglects the reflection that would occur even
in the absence of a barrier, so the conductance of the
actual device may be somewhat lower than the WKB estimate.)

Figure \ref{Fig:SB-variation}b shows the conductance
for different SB heights.
When the metal Fermi level falls in the middle of the
NT bandgap (as determined by the respective workfunctions
\cite{SBnote}),
the conductance at zero drain voltage shows a
symmetric dependence on the gate voltage.
There is significant conductance at gate voltages of 5-10~V,
but the ``turn-on" is very far from ideal,
with a conductance well below its maximum value of
$4e^2/h \approx 1.5 \times 10^{-4}$~S even at 20~V.

The underlying mechanism for the transistor action
is illustrated in Fig.~\ref{Fig:SB-variation}c.
Already at 4~V there is substantial carrier density
in the channel, but current is blocked by the SB.
Increasing the voltage difference between source and gate
electrodes leads to a large electric field
at the contact, reducing the width of the
Schottky barrier and allowing thermally-assisted tunneling.

If the metal Fermi level does not fall in the center
of the NT bandgap, the SB height is different
for electrons and holes, resulting in an asymmetric conductance curve
(dashed and dotted curve in Fig.~\ref{Fig:SB-variation}b).
Whenever the SB is large enough to effectively block current,
the device operates as a SB-FET.  For very small SBs,
however, the device operates as a normal (channel-limited) FET.

\begin{figure}
\begin{center}
\epsfig{file=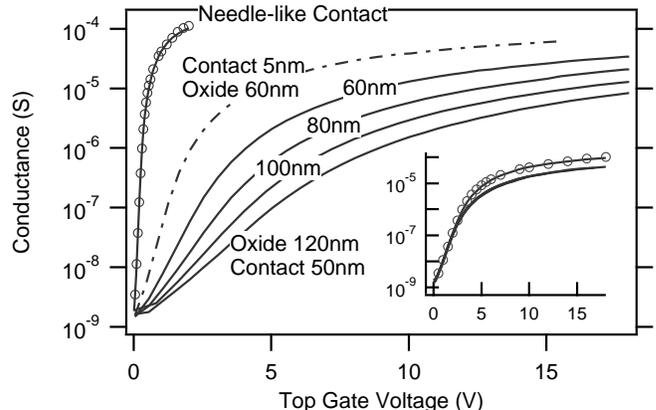,width=8.5cm,angle=0}
\caption{
\label{Fig:2D-Electrostatics}
Influence of the FET geometry on the device characteristics.
The four right-most curves correspond to
different thicknesses of the oxide
above the NT, as labeled, with all other parameters as in
Fig.~\ref{Fig:SB-variation}.
(The curve for 100~nm is the same as that in
Fig.~\ref{Fig:SB-variation}b for a SB height of 0.3~eV.)
The dot-dashed curve shows the
conductance when the contact thickness is reduced to 5~nm.
The curve at the left corresponds to a needle-like
metal electrode and cylindrical gate, see text.
Open circles are calculated as for the other curves,
solid curve uses the exact electrostatic kernel.
All calculations are at room temperature.
The inset shows the same curves, but for each curve
the gate voltage is rescaled by the voltage at which the
conductance is $10^{-8}$~S.
(The five rightmost curves cannot be distinguished on this scale,
forming a single line.)
}
\end{center}
\end{figure}

We now consider the crucial role of the device geometry.
We focus on the case of mid-gap barrier, where
the conductance is symmetric with respect to the gate voltage,
so it is sufficient to consider positive gate voltages.
(Our general conclusions apply equally to any case where the Fermi
level at the interface falls deep in the gap, so the barriers
for both electrons and holes are substantial.)
Figure~\ref{Fig:2D-Electrostatics} displays the conductance as
a function of gate voltage for a variety of device geometries.
As expected, the turn-on is better for thinner oxides.
However, even for the thinnest oxide (60~nm),
the device requires a rather large gate voltage for full turn-on.

In standard FETs, the conductance is controlled by
the electrostatic potential (and resulting carrier density)
in the channel;  so the key to improved performance is
increased capacitance between channel and gate by
reduction of the oxide thickness.
In contrast to this, for a SB-FET the performance is largely
controlled by the {\it electric field at the contact},
as illustrated in Fig.~\ref{Fig:SB-variation}c.
A sharper contact leads to field focusing, and hence
a larger field at the contact.
Thus,
%
the device performance can be dramatically improved
by varying the {\it geometry of the contacts},
in addition to the familiar dependence on gate distance.
For example,
if the thickness of the metal electrode is reduced from 50~nm to 5~nm,
there is a striking improvement of the device performance,
as shown in Fig.~\ref{Fig:2D-Electrostatics}.
The threshold voltage is lower,
and the conductance approaches the maximum value
at a lower gate voltages.
This improvement would not occur for channel-limited conduction.

The relative position of contact and gate is also important.
Most experiments reported to date have actually used
a bottom gate underneath a thick oxide (100 to 1000~nm).
From the field lines of Fig.~\ref{Fig:SB-variation}a,
it is clear that the electric field at the metal-NT contact
is larger when the gate is on the same side of the electrode
as the NT.
This suggests why, with a bottom gate,
substantial modulation of the conductance
was obtained even with thick oxides.

For good switching at a modest gate voltage,
one wishes to maximize the electric field at the contact
for a given gate voltage.
Thus the ideal geometry would be a sharp needle-like contact,
with a wrap-around cylindrical gate. The electrode
could be either a metal wire, or a metallic NT \cite{wrap-gate}.
The conductance for such a device
(with gate radius 50~nm and a SiO$_2$ dielectric)
is included in Fig.~\ref{Fig:2D-Electrostatics}.
The huge gain in performance is quite striking.

This simple geometry also provides an opportunity
to test our local approximation for the electrostatics of
the charge on the NT.
A solid line in Fig.~\ref{Fig:2D-Electrostatics} shows the
results using the exact kernel for this geometry \cite{Odintsov00}.
The difference is barely visible.
This justifies the approximation used for the
other geometries in Figs.~\ref{Fig:SB-variation}
and~\ref{Fig:2D-Electrostatics}.

The various device geometries lead to very different conductance
curves in Fig.~\ref{Fig:2D-Electrostatics}.
However, the difference is primarily in the scale of gate voltage
needed to modulate the conductance.
In fact, we can scale the five realistic devices to a single ``universal"
conductance characteristic, as demonstrated
in the inset of Fig.~\ref{Fig:2D-Electrostatics}.
Even the idealized needle-contact geometry is well approximated
by the same scaling function.
(The scaling is less closely obeyed at low temperature,
where the detailed shape of the barrier becomes more important.)
%
%
The scaling basically reflects the ratio of gate voltage
to electric field at the contact.
This scaling behavior is of great value in providing
a broad perspective on the performance of NT SB-FETs.

One puzzling aspect of NT devices has been the
dramatic effect of adsorbed gases on the behavior.
It has been naturally assumed that this effect arises
from doping of the NT~\cite{Jhi00,Collins00}.
However, recent work showed that the effect of adorbed
oxygen was qualitatively different than that of alkali
metals like potassium~\cite{Derycke02}, which are also
assumed to act as dopants.
\begin{figure}
\begin{center}
\epsfig{file=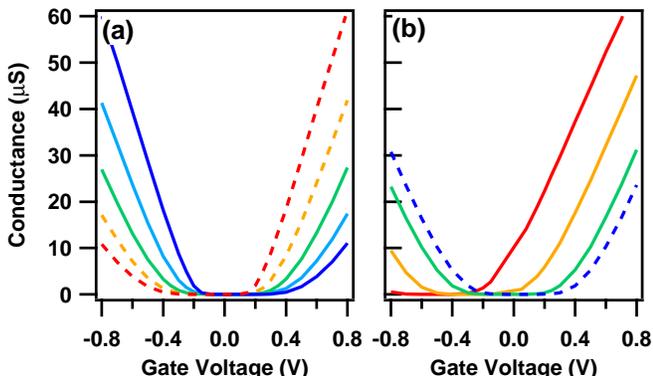,width=8.5cm,angle=0}
\label{Fig:Theory}
\caption{
Calculated conductance vs.\  gate voltage at room temperature,
varying
(a) the workfunction of the metal electrode, and
(b) doping of the NT.
In (a) the workfunction of the metal electrode is changed by
$-0.2$~eV (red dashed), $-0.1$~eV (orange dashed),
$0$~eV (green), $+0.1$~eV (light blue),
and $+0.2$~eV (blue), from left to right respectively.
In (b)
the doping atomic fraction is n-type
$10^{-3}$ (red), $5\times10^{-4}$ (orange),
and $10^{-4}$ (green), and p-type $10^{-4}$ (blue dashed),
from left to right, respectively.
(Colors refer to online version.)
}
\end{center}
\end{figure}

For NTs, the Schottky barrier is sensitive
to the metal workfunction \cite{Leonard00};
and workfunctions are well known to be sensitive to adsorbed gases
\cite{workfn}.
In most experiments, the metal contacts and NT have been directly
exposed to O$_2$.
Thus, one inevitable effect of gases will be to change
the metal workfunction.  (The NT workfunction may also be
changed; but because of the weaker binding to these
chemically inert structures, we focus on changes in
the workfunction of the metal electrode.)

Figure \ref{Fig:Theory}a shows the calculated conductance
vs.\  gate voltage, for varying workfunctions of the
metal source electrode.
(The gate workfunction is assumed to be unaffected,
since the bottom gate used in experiments is protected by the oxide.
A different gate workfunction would simply shift the zero of
gate voltage.)
Figure~\ref{Fig:Theory}b shows the same for varying doping of the NT.
For doping, the charge on the tube plays a central role,
so we use the idealized cylindrical geometry and exact
electrostatic kernel for these calculations.
We have already seen that the behavior for this model
is almost identical to that for realistic geometries,
aside from a scaling of the gate voltage;
so the results apply rather generally.

\begin{figure}
\begin{center}
\epsfig{file=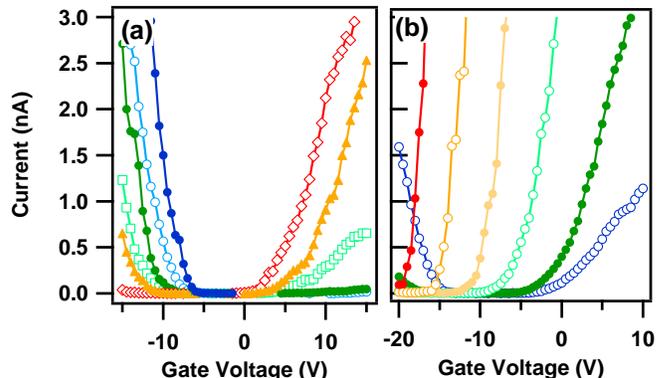,width=8.50cm,angle=0}
\label{Fig:Experiment}
\caption{
The experimentally measured effect of (a) oxygen adsorption and
(b) potassium doping on NT-FETs. In (a) the annealed (n-type) FET
has been exposed to oxygen for 2 minutes at
$P=0$~Torr (red), $P=10^{-4}$~Torr (orange),
$P=5\times 10^{-4}$~Torr (light green),
$P=5\times 10^{-3}$~Torr (dark green), $P=10^{-1}$~Torr (light blue),
and in ambient (blue), from left to right respectively.
Details are given in Ref.~[7]. In (b) the curves from
right to left (blue, dark and light green, light and dark orange, red)
correspond to increasing deposited amounts of potassium.
(Colors refer to online version.)
}
\end{center}
\end{figure}

The characteristics show a qualitatively different behavior
for doping vs.\  workfunction changes.
The workfunction difference
leads to a lowering of the conductance for one sign of gate voltage
and an increase for the opposite sign.
There is little change in the range of gate voltage
at which the FET is ``off", with conductance nearly zero.
For doping, in contrast,
the curves shift --- to negative gate voltages for n-type
doping, and to positive gate voltages for p-type doping.
At a sufficiently high level of doping,
a finite conductance is observed even at zero gate voltage.

This general behavior of the conductance is in striking
qualitative agreement with the experimental measurements
shown in Fig.~\ref{Fig:Experiment}.
(Details of the experiment are given elsewhere \cite{Derycke02}.)
We therefore conclude that the principle effect of
oxygen exposure in the experiment is not to dope the NT,
but rather to change the workfunction of the exposed portion
of the metal electrode.

\begin{figure}
\begin{center}
\epsfig{file=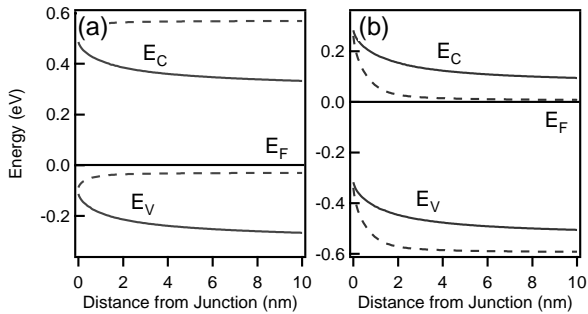,width=8.5cm,angle=0}
\caption{
\label{Fig:Bands}
Bending of NT valence and conduction bands at room temperature in the
case of (a) metal electrode workfunction increased by $0.2$~eV and
(b) n-type doping of the NT at atomic fraction $5\times10^{-4}$.
The solid lines corresponds to zero gate voltage. The dashed
line are for $-500$~mV in (a) and $+500$~mV in (b).
All energies are with respect to the Fermi energy.}
\end{center}
\end{figure}

The distinctive dependence on gate voltage for the two cases can be
understood from the respective band diagrams.
Figure \ref{Fig:Bands} shows the band diagrams,
for the case where the Fermi level originally falls at midgap.
In Figure \ref{Fig:Bands}a, the workfunction of the electrode
has been increased (e.g.\  by adsorbates) relative to the NT and gate.
For zero gate voltage there is no channel conductance,
and this is unaffected by the workfunction change.
For one sign of gate voltage, shown in Fig.~\ref{Fig:Bands}a,
when the channel turns on, the Schottky barrier is already reduced
by the workfunction change;  so the effect of the workfunction
change is to enhance the turn-on.
For the opposite gate voltage, when the channel turns on,
the Schottky barrier at the contact is increased by the
workfunction change, suppressing turn-on.
This gives the characteristic asymmetric turn-on,
without a shift in the voltage range where the
channel conductance is suppressed.

Doping, in contrast, shifts the bands in the channel,
and hence shifts the voltage range in which the channel is nonconducting.
This accounts for the most dramatic difference between
the two cases.
Figure \ref{Fig:Bands}b shows the behavior for n-type
doping. The width of the depletion barrier can be reduced by applying a
positive gate voltage, allowing electron tunneling through the thin barrier.
For negative gate voltage, however, it is extremely difficult
to turn on the FET conductance --- the gate voltage must be large
enough both to achieve inversion in the doped channel and
to narrow the Schottky barrier enough to permit tunneling.
Thus in the experiment of Fig.~\ref{Fig:Experiment}b,
turn-on at negative
gate voltages is seen only for the lowest doping levels studied.

In conclusion, the transistor action observed in
carbon nanotube FETs can be understood on the basis of transport
across a Schottky Barrier at the metal-NT contact.
The gate induces an electric field at the contact, which
controls the width of the barrier and hence the current.
A sharper contact leads to focusing of the electrical field,
allowing operation at lower gate voltages.
Changes in workfunction, e.g.\  by adsorbed gases, affect the Schottky barrier
and hence the device characteristics.
By comparing calculations with experimental data for
FETs exposed to oxygen or doped with potassium,
we suggest that the main effect of oxygen exposure is to change the workfunction of
the metal contact rather than to dope the NT.

S.\ H.\ thanks the Deutsche Forschungsgemeinschaft for financial support
under the Grant number HE3292/2-1.

\end{multicols}

\begin{references}
\bibitem{local} We take the contribution to the electrostatic
potential at any point on the NT from charge on the NT to be
a function only of the charge at that point.
The function is determined numerically for a translationally
invariant geometry (where the approximation becomes exact).
\bibitem{SBnote} In general there may be ``pinning" of the lineup at
the metal-NT interface; but here we neglect such pinning,
because it is much less important for NT Schottky barriers
than for planar interfaces \cite{Leonard00}.
\bibitem{wrap-gate} The latter would be a particularly elegant approach,
and such a junction was considered by Odintsov \cite{Odintsov00}.
Here we focus on the case of a metallic wire having the same radius
as the NT;
but we find little difference in performance between the two cases.
\end{references}
\end{document}